\documentclass[aps,amssymb,amsmath,prl,reprint,superscriptaddress,noshowpacs]{revtex4-1}

 \fussy

\usepackage{times}
\usepackage[T1]{fontenc}
\usepackage[latin1]{inputenc}
\usepackage[english]{babel}

\usepackage{textcomp}
\usepackage{bm}

\tighten

\usepackage[pdftex]{graphicx}
\usepackage[usenames]{color}

\newcommand{\figwidth}{0.45\textwidth} 

\newcommand{\vect}[1]{\mathbf{#1}}

\newcommand{\affil}{Center for Nanophotonics, FOM Institute for Atomic and Molecular Physics, Science Park 104, 1098 XG Amsterdam, The Netherlands}

\begin{document}
\author{Martin Frimmer}\email{frimmer@amolf.nl}\homepage{http://www.amolf.nl/}
\affiliation{\affil}
\author{Toon Coenen}
\affiliation{\affil}
\author{A. Femius Koenderink}
\affiliation{\affil}

\title{Signature of a Fano-resonance in a plasmonic meta-molecule's local density of optical states}

\begin{abstract}
We present measurements on plasmonic meta-molecules under local excitation using cathodoluminescence which show a spatial redistribution of the local density of optical states (LDOS) at the same frequency where a sharp spectral Fano-feature in the extinction cross section has been observed. Our analytical model shows that both near- and far-field effects arise due to interference of the same two eigenmodes of the system. We present quantitative insights both in a bare state, and in a dressed state picture that describe plasmonic Fano interference either as near-field amplitude transfer between three coupled bare states, or as interference of two uncoupled eigenmodes in the far field. We identify the same eigenmode causing a dip in extinction to strongly enhance the radiative LDOS, making it a promising candidate for spontaneous emission control.
\end{abstract}

\date{Submitted: September 25, 2011}

\maketitle 

\paragraph{Introduction.}
Interference is ubiquitous in physics. Significant recent advances in optics as well as  quantum physics hinge on interference, inherent in the wave nature of light and matter, and the superposition principle. In quantum optics, the Fano-effect and its occurrence in electromagnetically induced transparency (EIT) have in particular triggered tremendous interest as phenomena relying on quantum interference~\cite{Fleischhauer2005} in light-matter coupling. In  EIT, a strongly absorbing atomic vapor coupled to an intense pump field acquires a narrow transparency window, with unusual features, such as ultralow group velocities and huge nonlinearities.
These extraordinary properties have attracted the interest of the field of nano-photonics, the science of engineering the generation, the propagation and the absorption of light on a subwavelength scale~\cite{Novotny2006}. The aspiration of nano-optical circuitry with powerful functionality led to the development of optical meta-materials. These artificial materials are composed of meta-atoms, designed building blocks giving rise to peculiar properties not found in natural materials~\cite{Soukoulis2011}. Inspired by quantum optics, scientists have identified plasmonic meta-molecules whose optical properties mimic EIT-lineshapes in atomic vapors, an effect termed `plasmon-induced transparency' (PIT)~\cite{Zhang2008}, based on the Fano-interference of a super- and a sub-radiant mode.
Even without the benefit of a full electrodynamic model reaching beyond brute force numerical simulations, remarkable intuition and simple electrostatic arguments have led to the development of several structures exhibiting PIT~\cite{Luk'yanchuk2010,Liu2009a,Verellen2009,Zhang2008,Fan2010,Hao2009,Hentschel2011,Hentschel2010,Lassiter2010,Alonso-Gonzalez2011}.
While in PIT plasmonic meta-molecules control the propagation of light by creating narrow dark resonances useful for slow light or sensing,  another class of nano-structures termed `optical antennas' is currently being developed to tailor light matter interaction~\cite{Kuehn2006,Anger2006,Kinkhabwala2009}.  Antennas exploit bright resonances to enhance the emission of light.
Practically all aspects of spontaneous emission control by optical antennas rely on designed enhancement of the local density of optical states (LDOS), arguably the most fundamental quantity in nano-optics~\cite{Novotny2006}.
An outstanding question is what the LDOS of PIT structures is, and if one can use the narrow dark lines of PIT to improve optical antennas. To answer this question it is essential to unravel which modes are involved in PIT, how they project on localized driving, and how they give rise to our observable, i.e. far field radiation.

In this Letter, we map the LDOS of a plasmonic molecule known to exhibit PIT. Under near-field driving using cathodoluminescence (CL) we observe a marked spatial redistribution of LDOS occurring at wavelengths coincident with the PIT-dip in extinction, thereby relating far-field data to the near-field LDOS.
We present a model that identifies the diagonal `dressed states' of the meta-molecule and proves the LDOS feature to be due to interference of the same eigen-states that are responsible for PIT in plane wave extinction. We draw three remarkable conclusions beyond the analogy with atomic EIT. Firstly, we find a `screening' state that significantly renormalizes the interaction. Secondly, we conclude that all observations can be viewed purely as far field interference of two eigen-modes of the system. Thirdly and surprisingly, we identify the sub-radiant mode to be a promising candidate to enhance brightness and rate of spontaneous emission into the far field.
\begin{figure}
\includegraphics[width=\figwidth]{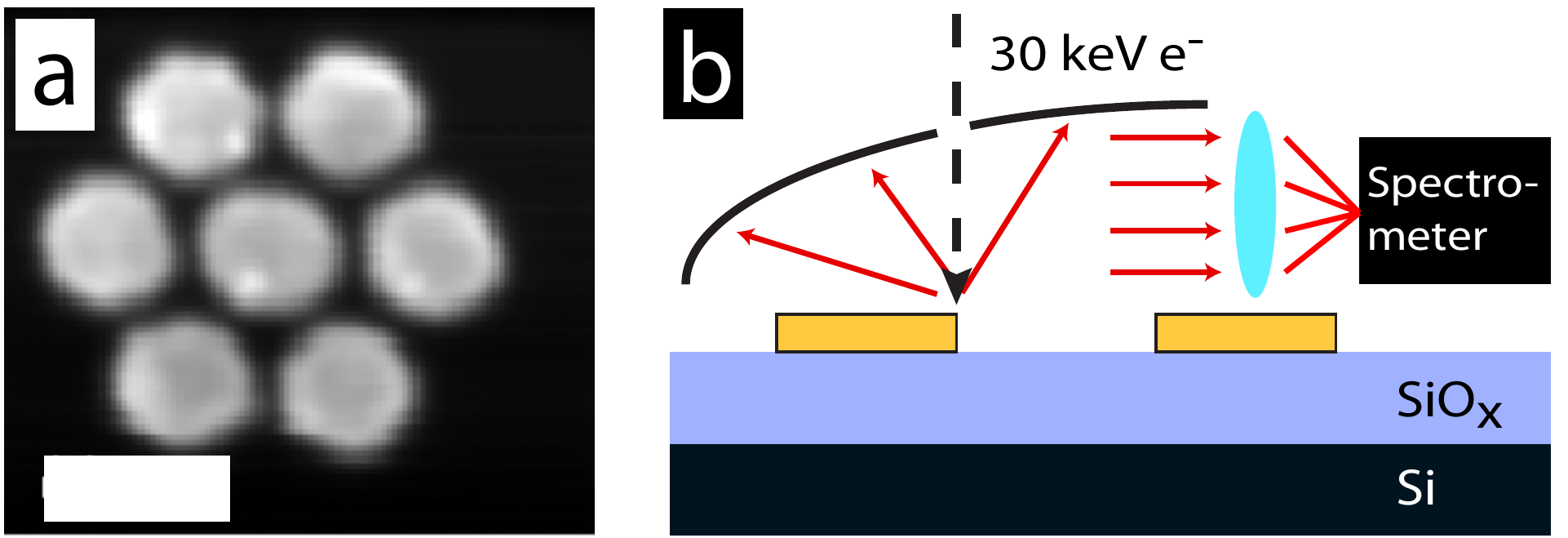}
\caption{(a) SEM micrograph of plasmonic heptamer. Scale bar is 200\,nm. (b) Sketch of experimental setup showing sample, impinging electron beam, parabolic mirror and optics guiding CL to a  spectrometer. }
\label{SetupModes}
\end{figure}

\paragraph{Experimental.}
\begin{figure}
\includegraphics[width=\figwidth]{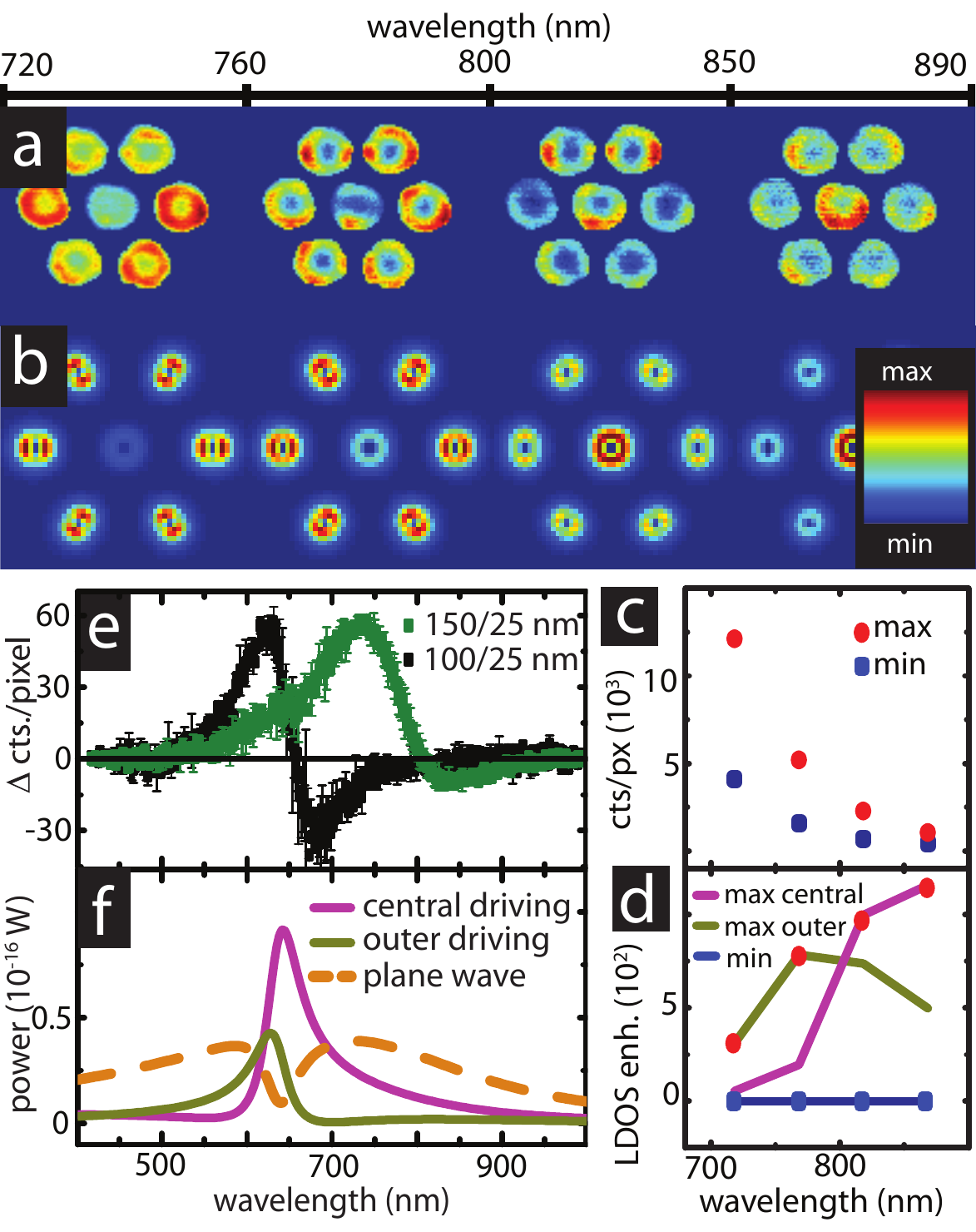}
\caption{(a) Measured CL excitability maps of plasmonic heptamer. (b) Calculated maps of radiative LDOS. Insert: colorbar for (a) and (b). (c) Limits of colorscale for (a). Background around 700\,nm stems from oxide substrate. (d) Calculated radiative LDOS enhancement on central (purple) and outer (olive) particles of heptamer with particle diameter/gapwidth 150/25\,nm. Blue and red symbols denote colorscale limits of frames in (b). (e) Measured differential excitability spectra showing intensity difference between outer and central particles. Green (black): Particle diameter/gapwidth 150/25\,nm (100/25\,nm). Errorbars denote standard deviation of 4 (2) structures. (f) Calculated power radiated by super- and subradiant eigenmodes of heptamer upon driving central (purple)/outer (olive)/all (orange dashed) particles with vertical electric field.}
\label{CLmaps}
\end{figure}
We fabricated plasmonic heptamers on  a Si wafer covered with 1\,\textmu m thermal oxide  by electron beam lithography~\cite{Sersic2011}, thermal evaporation of 35\,nm Au, and lift off. Each heptamer consists of nominally identical particles arranged on the corners and center of a  hexagon. A SEM image of a typical heptamer with particle diameter 150\,nm, and gap width 25\,nm is shown in Fig.~\ref{SetupModes}(a).
The CL measurements were performed in a scanning electron microscope, sketched in Fig.~\ref{SetupModes}(b).
CL maps are acquired by raster scanning the focused electron beam (30\,keV, waist <\,5\,nm) across the sample. Emitted light collected with a parabolic mirror (acceptance angle 4.6 sr) is spectrally analyzed on a silicon CCD camera \cite{[{For details of the experimental setup see }] Coenen2011}. The signal hence represents `CL excitability'  as a function of detection wavelength and spatial excitation coordinate. In Fig.~\ref{CLmaps}(a) we show  CL collected from the heptamer in Fig.~\ref{SetupModes}(a)  as a series of spatial excitation maps by binning the data into 50\,nm wavelength slices. We identified pixels on substrate \emph{vs.} on Au particles by thresholding the SEM data collected in parallel with the CL. We clamp the color value for all substrate pixels to the smallest value obtained on any Au particle to maximize color contrast for the regions of interest we analyze, \emph{i.e.} the particles. The minimum and maximum values of the colormaps in Fig.~\ref{CLmaps}(a,b) [colorbar in insert in Fig.~\ref{CLmaps}(b)] are plotted in Fig.~\ref{CLmaps}(c,d), respectively.
While at wavelengths shorter than 800\,nm in Fig.~\ref{CLmaps}(a) the outer particles appear brighter, \emph{i.e.} their excitability is higher, the situation is reversed at wavelengths longer than 800\,nm, where the inner particle is more excitable. Local hotspots are due to surface roughness inherent to the fabrication.
\begin{figure}
\includegraphics[width=\figwidth]{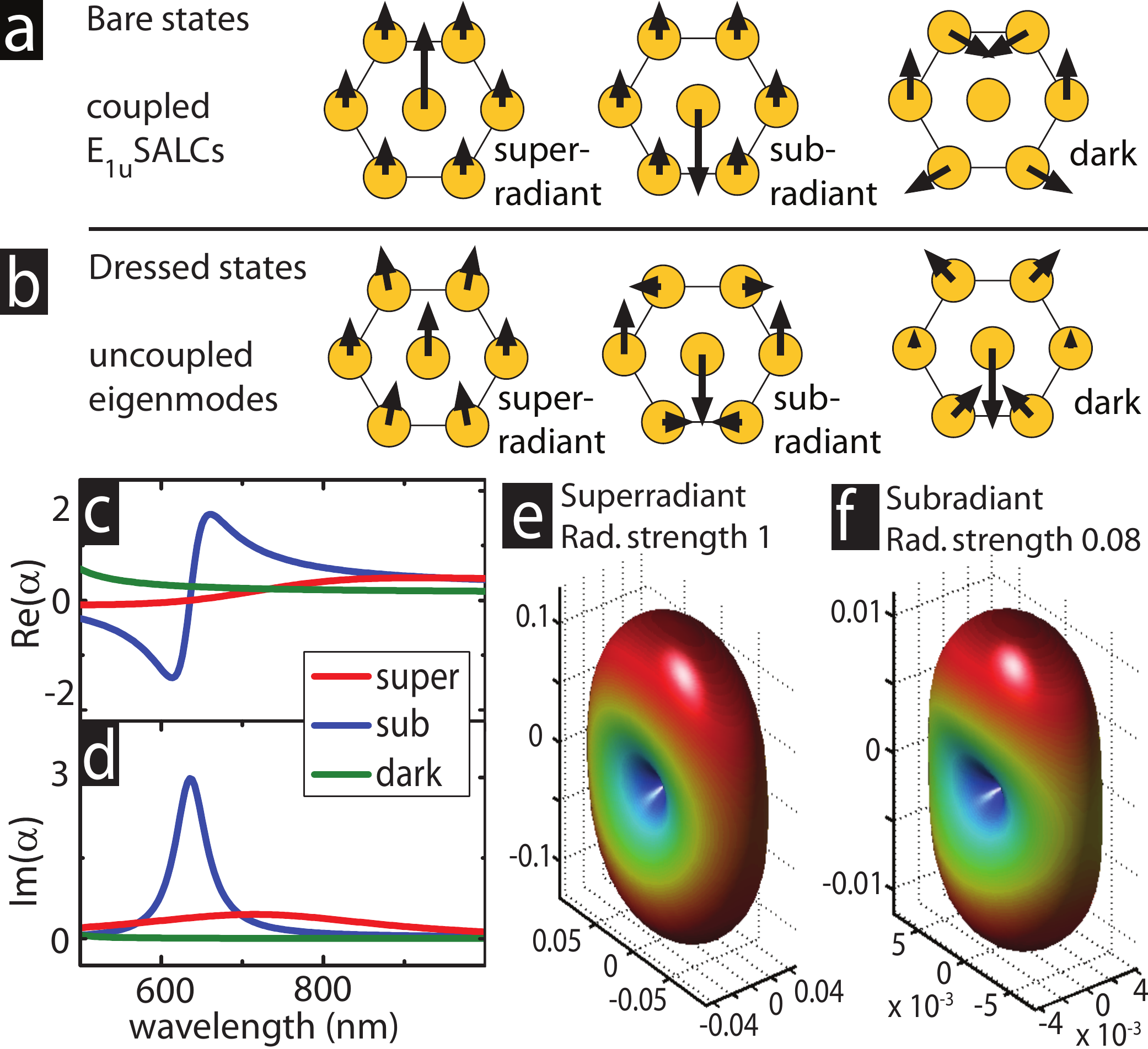}
\caption{(a) Bare state  non-diagonal polarization  basis, consisting of a super-radiant, sub-radiant and dark  symmetry adapted linear combination (SALC) of E$_\text{1u}$ derived from group theory. (b) Dressed state basis of decoupled superradiant, subradiant and dark eigenmodes (small out-of-phase components  not visualized).
(c) Real and (d) imaginary parts of polarizabilities of the eigenmodes in $10^{-31}\,$Cm$^2$/V. (e) and (f) are radiation patterns of superradiant and subradiant eigenmodes in W/sr. Radiative strength indicates total radiated power relative to superradiant eigenmode.}
\label{Fig_RadPatt}
\end{figure}
To quantify the observed swap in local excitability we discriminate pixels belonging to outer or inner particles, spatially average their spectra and subtract them from each other. The resulting `differential excitability' is therefore positive if the outer particles are more excitable than the inner one. This procedure eliminates the broad background fluorescence of the oxide layer around 650\,nm, which causes the overall increase in absolute signal towards the blue in Fig.~\ref{CLmaps}(c). Note that this background together with the vanishing detector efficiency towards the infrared cause the apparent discrepancy of Fig.~\ref{CLmaps}(c) and (d).  Figure~\ref{CLmaps}(e) shows differential excitability spectra for heptamers with particle diameter/gap size 150/25($\pm$5)\,nm averaged over four structures [green squares] and 100/25($\pm$5)\,nm, averaged over two [black squares].
The wavelength where the differential excitability changes sign, \emph{i.e.} where excitability swaps from outer to inner particles, appears around 800\,nm for the large heptamers, as reported  in Fig.~\ref{CLmaps}(a), and blue-shifts to 650\,nm for smaller particles.

\paragraph{Electrodynamic model.}
We model the plasmonic heptamers with a fully electrodynamic coupled dipole model~\cite{Koenderink2006}. First, we show that this model captures both the far-field Fano interference, and the excitability swap in our experiment. Subsequently, we derive new insights by identifying the bare and dressed states in analogy to the atomic system. In our model, the response of each particle to an electric field, \emph{i.e.} its dipole moment $\vect{p}={\bm \alpha} \mathbf{E}$, is proportional to its Lorentzian polarizability tensor ${\bm \alpha}$~\footnote{We model the particles as oblate spheroids with short axis 15\,nm and long axis 50\,nm [75\,nm for LDOS maps in Fig.~\ref{CLmaps}(b)] in air~\cite{Bohren1983} and extend the static polarizability by a radiation damping term~\cite{Koenderink2006}. For $\epsilon$ we assume a Drude model ($\omega_0=4.76\cdot10^{15}$\,rad\,s$^{-1}$, $\gamma=8.5\cdot10^{13}$\,s$^{-1}$)}. The resulting set of coupled linear equations can be written in the form $\vect{p}=\vect{M^{-1}}\cdot \vect{E}_{ext}$,
where $\vect{E}_{ext}$ is the incident field driving each particle.
Both $\vect{E}_{ext}$ and $\vect{p}$ have 3$N$=21 elements for $N$=7 particles. The interaction matrix ${\bf M}$ has the inverse of ${\bm \alpha}$  on the diagonal, while the off-diagonal elements describe the interaction between dipoles as set by the electrodynamic Green's function.
For any dipole assembly $\vect{M}$ can be inverted to find the polarization state $\vect{p}$ induced by any driving $\vect{E}_{ext}$. Since this linear problem contains full electrodynamic interactions, its solution allows to calculate any near- and far-field observable. Specifically, the intensity of generated CL is proportional to the component of the radiative LDOS along the impinging electron beam~\cite{Garc'iadeAbajo2010,Kuttge2009}.

\begin{figure}
\includegraphics[width=\figwidth]{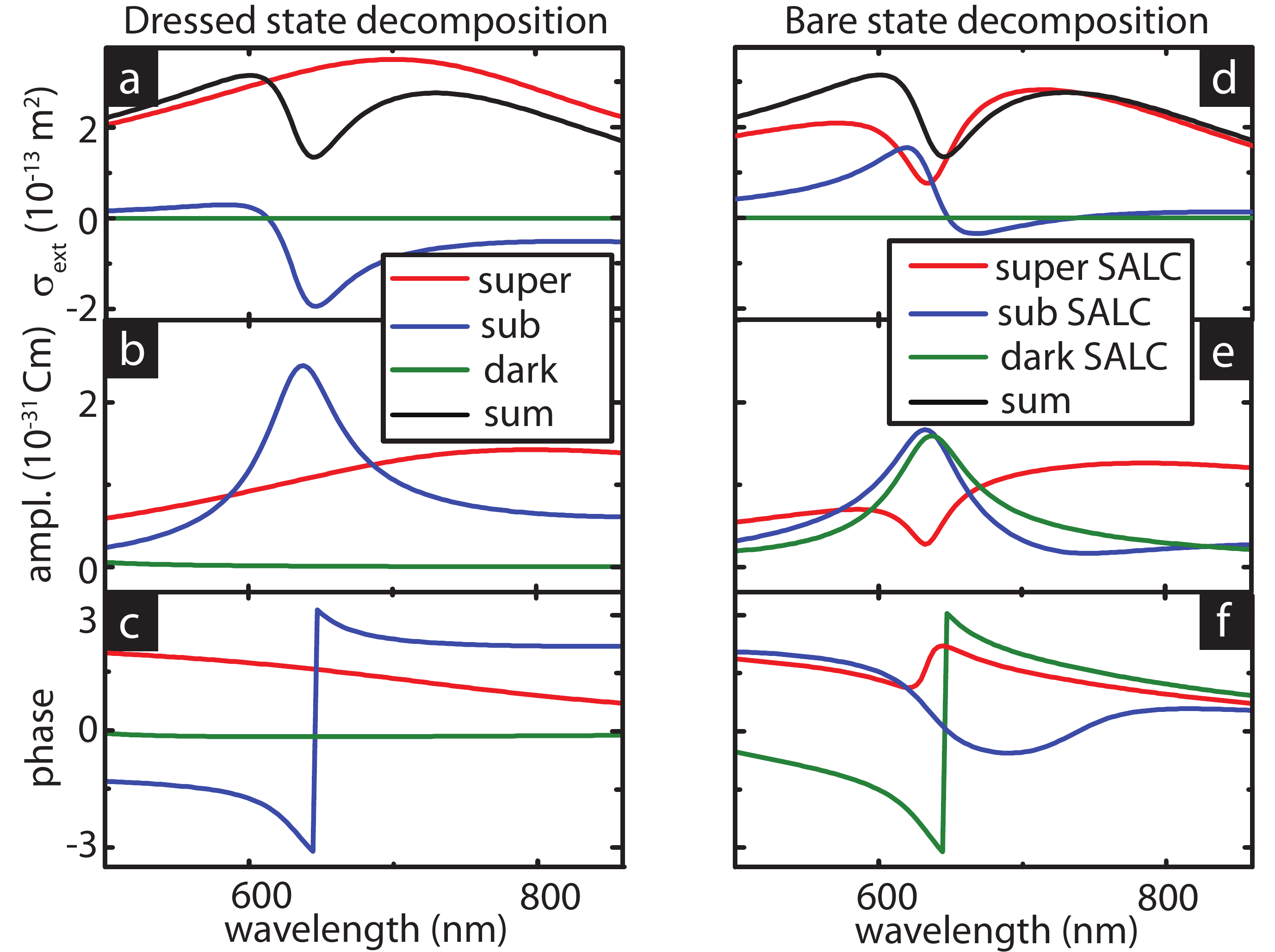}
\caption{(a) Calculated contribution of eigenmodes [Fig.~\ref{Fig_RadPatt}b)] to extinction cross section of plasmonic heptamers. (b) Eigenmode amplitude and (c) phase relative to driving field. (d,e,f) Corresponding graphs in the  non-diagonal bare state basis [Fig.~\ref{Fig_RadPatt}a)].}
\label{Fig_Amps}
\end{figure}

Figure~\ref{CLmaps}(b) shows calculated maps of the radiative LDOS for driving perpendicular to and located 40\,nm above the plane of the meta-molecule. As in the experimental data, the ring-shaped profiles per particle imply that we  detect in-plane induced particle polarizations, despite the out-of-plane incident electron beam~\cite{Yamamoto2011}. The calculations show the measured swapping of LDOS from outer to inner particles with increasing wavelength [Fig.\ref{CLmaps}(b)] as well as the shift of spectral features to shorter wavelengths with decreasing particle diameters [Fig.~\ref{CLmaps}(f)].
As we show below, the Fano-dip in extinction~\cite{Hentschel2010} calculated using the same model [see Fig.~\ref{Fig_Amps}] coincides with the spectral position of the excitability swap in CL, pointing at a direct relation between LDOS and Fano interference.

\paragraph{Symmetry of the heptamer and eigenmodes.}
While it is gratifying that our model quantitatively confirms both the Fano extinction dip and the concomitant CL signature, the analysis sofar provides little insight.
%
Three essential steps simplify the problem. Firstly, as pointed out by Mirin \emph{et al.}~\cite{Mirin2009}, the heptamer bears D$_{\text{6h}}$ symmetry. Therefore, Symmetry Adapted Linear Combinations (SALCs) of dipole moments~\cite{Wilson1934,*Rosenthal1936} cast the electrodynamic coupling matrix $\vect{M}$ into block diagonal form since group theory usefully extends beyond electrostatic hybridization~\cite{Prodan2003}, allowing a symmetry-based decoupling also for calculations of scattering.
Secondly, since we rely on far field detection, we focus our attention on the infrared-active E$_{\text{1u}}$ irreducible representations. Due to the degeneracy of horizontal and vertical polarization in the sixfold symmetry we are left with a three-dimensional subspace. Our choice of SALCs for E$_{\text{1u}}$ (vertical polarization only) are shown in Fig.~\ref{Fig_RadPatt}(a). Two SALCs are the in and out of phase superpositions of one hexamer and the single particle E$_{\text{1u}}$ modes pointed out in~\cite{Mirin2009}. In addition to the superradiant SALC (large net dipole moment) and  subradiant SALC (threefold smaller dipole moment), symmetry requires a third dark SALC with zero net dipole moment.
As a third essential step we note that while symmetry decouples different irreducible representations, we must still diagonalize the E$_{\text{1u}}$ submatrix  of $\vect{M}$ to find the true decoupled eigenmodes.

Remarkably, the three eigenvectors of E$_{\text{1u}}$, sketched in Fig.~\ref{Fig_RadPatt}(b), are almost unchanged across the frequency range from 400 to 1000\,nm. The complex eigenvalues show strong dispersion, as shown in Fig.~\ref{Fig_RadPatt}(c,d). These eigenvalues are the `eigenpolarizabilities' of the eigenmodes under which the E$_{\text{1u}}$ submatrix of $\vect{M}$ is diagonal and which therefore are by definition decoupled. Considering the dipole distributions in Fig.~\ref{Fig_RadPatt}(b) we can classify the first  eigenmode as superradiant, therefore featuring a broad eigenpolarizability [Fig.~\ref{Fig_RadPatt}(c,d), red line], reminiscent of the superradiant SALC in that all dipole moments are approximately aligned [compare Fig.~\ref{Fig_RadPatt}(a,b)]. The second eigenmode is subradiant with a narrow resonant eigenpolarizability  around 630\,nm, i.e. at the observed Fano dip and LDOS feature [Fig.~\ref{Fig_RadPatt}(c,d), blue].
Due to the coupling set by $\vect{M}$, this mode is remarkably different from the sub-radiant SALC.
The third mode has a narrow resonance at significantly shorter wavelengths, beyond our range of interest [green]. Since its eigenpolarizability is negligible around 630\,nm, the Fano interference is explained in terms of just the super- and subradiant modes [red and blue in Fig.~\ref{Fig_RadPatt}(c,d)]. These modes have largely overlapping radiation patterns [Fig.~\ref{Fig_RadPatt}(e,f)], although with a twelvefold ratio in integrated radiated intensity at identical amplitude. While the eigenmodes are by definition decoupled, the excellent radiation pattern overlap implies that nearly completely destructive or constructive interference can occur in the far field.

\paragraph{Two eigenmodes explain all observations.}
To demonstrate that two eigenmodes capture all the physics observed both under local and plane wave driving, we perform calculations using just the super- and subradiant eigenmodes. The purple curve in Fig.~\ref{CLmaps}(f) shows the radiated power when only the central particle is driven.  The total radiated power (proportional to CL intensity generated on the central particle)  shows an asymmetric peak at 650\,nm with a steep slope on its blue side, and a strongly broadened wing on its red side.  The localized driving mainly projects on the resonant subradiant mode. The superradiant mode provides a weak contribution that interferes destructively on the blue, and constructively on the red side of the resonance, leading to the typical asymmetric Fano-lineshape. Importantly, when driving the outer particle [olive curve in Fig.~\ref{CLmaps}(f)], the peak asymmetry is reversed, as the relative phase between excitation of the broad superradiant and narrow subradiant mode is swapped. This asymmetric broadening in opposite directions underlies the measured dispersive differential excitability in Fig.~\ref{CLmaps}(e), and signifies interference of the two decoupled eigenmodes on the detector.  When driving the heptamer by a plane wave [Fig.~\ref{CLmaps}(f), orange dashed line], we  find that the same two modes cause the Fano-dip reported in literature~\cite{Hentschel2010}.
The dark Fano-feature in extinction coincides spectrally with the asymmetric CL peaks, i.e. the asymmetric radiative LDOS enhancement, evident in local excitation.
In contrast to local excitation, plane wave driving strongly drives the superradiant mode, causing the weakly excited subradiant eigenmode to appear as a narrow dip on a high background. We conclude that the measured redistribution of LDOS in CL data  signifies the same interference mechanism as the Fano dip in extinction, though with very different superradiant and subradiant mode amplitudes.

\paragraph{Far-field interference in eigen-basis.}
To gain further insight in the response of the individual modes, we focus on plane-wave driving and Fano-interference in the extinction cross-section $\sigma_{ext}$ of the heptamer [black line Fig.~\ref{Fig_Amps}(a)].
In Fig.~\ref{Fig_Amps}(b,c) we plot the induced (complex) polarization $\vect{p}$ of the eigenmodes. Extinction (work done by the driving $ \text{Im} [\vect{E}_{ext}\cdot \vect{p}]$, normalized to incident intensity)  can readily be split into contributions from different modes [Fig.~\ref{Fig_Amps}(a)]. The total extinction cross-section is in excellent quantitative agreement with reported numerical results~\cite{Hentschel2010}, underlining the suitability of a dipole model. The superradiant mode provides a broad positive extinction [red line in Fig.~\ref{Fig_Amps}(a)].  The Fano-dip in the sum is created by the subradiant mode [blue]. Its surprising negative contribution to $\sigma_{ext}$  indicates that the mode feeds energy back into the driving field, equivalent to destructive far-field interference. This energy cannot result from direct driving of the superradiant mode and subsequent amplitude transfer~\cite{Alzar2002}, since the eigenmodes are strictly decoupled.  Since the projection of plane wave driving on the eigenmodes has no frequency dependence, the excitation amplitudes in Fig.~\ref{Fig_Amps}(b) simply follow the eigenpolarizabilities in Fig.~\ref{Fig_RadPatt}(c,d). On the Fano-dip, the subradiant mode's net dipole moment compensates that of the superradiant mode such that their superposition barely radiates. Importantly, the subradiant eigenmode must possess nonzero dipole moment (and excellent radiation pattern overlap with the superradiant mode) to lead to a PIT dip in extinction. It is this nonzero dipole moment together with a very strong, narrow eigen-polarizability [blue lines in Fig.~\ref{Fig_RadPatt}(c,d)] that also ensures that the subradiant mode can give rise to an enhanced radiative LDOS at the central particle, as seen from Fig.~\ref{CLmaps}(f). This eigenpolarizability exceeds the one of a single isolated particle and is narrower in width.
We therefore draw the counterintuitive conclusion that introducing sub-radiant modes that are usually associated with dark PIT phenomena can actually enhance the capability of optical antennas to  create bright and efficient emitters with a large radiative LDOS. The theoretical analysis tool that we have applied to the specific case of a heptamer will be of great value to explore the application of PIT to spontaneous emission enhancement. Our approach goes beyond both brute force numerical techniques and electrostatic hybridization and provides a generic, analytical framework that for the first time reveals the true eigenmodes of a structure exhibiting PIT, their polarizability, brightness, and response to any near- or far-field driving.

\paragraph{Identification of bare and dressed states.}
Fano interference is inextricably linked to coherent transfer of amplitude to states that are not directly driven~\cite{Alzar2002}. Indeed, this is the picture in which the discussion of PIT has been led thus far~\cite{Luk'yanchuk2010,Liu2009a,Verellen2009,Zhang2008,Fan2010,Hao2009,Hentschel2011,Hentschel2010,Lassiter2010,Alonso-Gonzalez2011}.
Necessarily, the eigenmode basis of any linear system never involves amplitude transfer, as eigenmodes are decoupled. Figure~\ref{Fig_Amps}(a-c) confirms this viewpoint, which is analogous to analyzing atomic EIT in terms of dressed states~\cite{Fleischhauer2005} that decay to the same continuum with opposite phase. The alternative view on EIT, in which amplitude transfer \emph{does} occur, is as an interference of different pathways of coherently coupled bare states~\cite{Fleischhauer2005}. The basis of SALCs [Fig.~\ref{Fig_RadPatt}a] provides the analog of such bare states in EIT for the PIT system. The contributions of the three SALCs to $\sigma_{ext}$ plotted in Fig.~\ref{Fig_Amps}(d) reveal a dip in the broad band of the super-radiant SALC [red line]. However, also the sub-radiant SALC [blue] contributes significantly.
Even though the contribution of the dark SALC [green] to $\sigma_{ext}$ is strictly zero it is crucial for the Fano-dip. Figure~\ref{Fig_Amps}(e) shows that the dark SALC acquires an amplitude at the Fano-dip as large as that of the subradiant SALC.
However, around the Fano dip the polarization $\vect{p}$ of the dark SALC is locked in amplitude and phase to the subradiant SALC [Fig.~\ref{Fig_Amps}(e,f)]. One can hence view the dark SALC as a screening effect that allows to describe the three-state interaction as the interaction between just two oscillators~\cite{Alzar2002}, with renormalized resonance frequencies and strengths. In conclusion, when taking the SALCs as a basis of bare states in which $\vect{M}$ is not diagonal, amplitude transfer occurs to a linear combination of the sub-radiant and dark SALC [Fig.~\ref{Fig_Amps}(e)].
The link between the complementary interpretations of EIT in scattering systems via either amplitude transfer or far-field interference is ultimately provided by the optical theorem which constrains $\vect{M}$ and thereby inextricably links radiated power and induced complex dipole moments to satisfy energy conservation.

\paragraph{Conclusions.}
In conclusion, we measured a spatial redistribution of the LDOS of plasmonic heptamers using CL at the spectral position that coincides with the reported Fano extinction dip. Both LDOS redistribution and Fano dip result from interference in the far-field of the same two eigenmodes, excited in different coherent superpositions. Our findings bear a plethora of exciting prospects to harness the near field of plasmonic molecules, especially in the context of spontaneous emission, or control of any process in sensing, spectroscopy and non-linear optics that benefits from enhanced LDOS. Strikingly,  optimizing a Fano dip in extinction requires engineering of radiation pattern overlap between two involved modes that ensures an enhanced LDOS by constructive interference. Also, completely dark SALCs may be utilized to optimize resonances and interaction strengths, similar to the screening dark SALC in the heptamer. For fluorescence applications, one typically requires simultaneous optimization of pump field generated from a far field beam, and optimization of the LDOS at the redshifted emission frequency. Our generic quantitative solution approach may yield universal bounds on how to optimize the solution to such a problem using optical antennas with a Fano resonance.
%
%
\begin{acknowledgments}
We thank Peter Nordlander for fruitful discussions.
This work is part of the research program of the ``Stichting voor Fundamenteel
Onderzoek der Materie (FOM)'', which is financially supported by the
``Nederlandse Organisatie voor Wetenschappelijk Onderzoek (NWO)''. Toon Coenen is financially supported by Nano-NextNL, a nanotechnology program funded by the Dutch ministry of economic affairs.
\end{acknowledgments}

\bibliography{FrimmerBib}

\end{document}